\documentclass[letterpaper]{article} 
\usepackage{aaai25}  
\usepackage{times}  
\usepackage{helvet}  
\usepackage{courier}  
\usepackage[hyphens]{url}  
\usepackage{graphicx} 
\usepackage{natbib}  
\usepackage{caption} 
\usepackage{algorithm}

\usepackage{booktabs}
\usepackage{amssymb}
\usepackage{breqn}
\usepackage[utf8]{inputenc}
\usepackage{algpseudocode}
\algnewcommand\algorithmicinput{\textbf{INPUT:}}

\usepackage{newfloat}
\usepackage{listings}
\DeclareCaptionStyle{ruled}{labelfont=normalfont,labelsep=colon,strut=off} 
\floatstyle{ruled}
\newfloat{listing}{tb}{lst}{}
\floatname{listing}{Listing}
\title{AlphaForge: A Framework to Mine and Dynamically Combine Formulaic Alpha Factors  }
\author{
    Hao Shi\textsuperscript{\rm 1}, 
    Weili Song\textsuperscript{\rm 2}, 
    Xinting Zhang\textsuperscript{\rm 1},
    Jiahe Shi\textsuperscript{\rm 3}, 
    Cuicui Luo\textsuperscript{\rm 1}\thanks{Corresponding author.},
    Xiang Ao\textsuperscript{\rm 5}, 
    Hamid Arian\textsuperscript{\rm 6}, 
    Luis Seco\textsuperscript{\rm 7}
}
\affiliations {
    \textsuperscript{\rm 1}University of the Chinese Academy of Sciences, Beijing, China \\
    \textsuperscript{\rm 2}Renaissance Era Investment Management Co., Ltd, Beijing, China \\
    \textsuperscript{\rm 3}Shangqiu Normal University, Shangqiu, China \\
    \textsuperscript{\rm 5}Institute of Computing Technology, University of Chinese Academy of Sciences, Beijing, China \\
    \textsuperscript{\rm 6}York University, Toronto, Canada \\
    \textsuperscript{\rm 7}University of Toronto, Toronto, Canada \\
    \{shihao22@mails.ucas.ac.cn, weilisong@hnu.edu.cn, 850446810@qq.com, s2021371@siswa.um.edu.my, luocuicui@ucas.ac.cn, aoxiang@ict.ac.cn, harian@yorku.ca, luis.seco@utoronto.ca\}
}

\usepackage{bibentry}

\begin{document}

\maketitle

\begin{abstract}
The complexity of financial data, characterized by its variability and low signal-to-noise ratio, necessitates advanced methods in quantitative investment that prioritize both performance and interpretability.Transitioning from early manual extraction to genetic programming, the most advanced approach in the alpha factor mining domain currently employs reinforcement learning to mine a set of combination factors with fixed weights.  However, the performance of resultant alpha factors exhibits inconsistency, and the inflexibility of fixed factor weights proves insufficient in adapting to the dynamic nature of financial markets. To address this issue, this paper proposes a two-stage formulaic alpha generating framework AlphaForge, for alpha factor mining and factor combination. This framework employs a generative-predictive neural network to generate factors, leveraging the robust spatial exploration capabilities inherent in deep learning while concurrently preserving diversity. The combination model within the framework incorporates the temporal performance of factors for selection and dynamically adjusts the weights assigned to each component alpha factor. Experiments conducted on real-world datasets demonstrate that our proposed model  outperforms contemporary benchmarks in formulaic alpha factor mining. Furthermore, our model exhibits a notable enhancement in portfolio returns within the realm of quantitative investment and real money investment.
\end{abstract}

%

\section{Introduction}
A central challenge in the field of quantitative investment is stock trend forecasting. This difficulty arises from the inherent characteristics of stock data, featuring a low signal-to-noise ratio and considerable noise \cite{y14}. Professionals and researchers commonly employ a strategy involving the extraction of Alpha factors from raw historical data to predict future returns.
The contemporary approaches to Alpha factor mining can be broadly categorized into two main methodologies: machine learning methods and formulaic Alpha methods. Deep learning models such as LSTM\cite{y5} and Historical Inference Sequence Transformers (HIST)\cite{y23},  among others, are commonly employed to generate more complex Alphas, which often lack interpret-ability. On the other hand, formulaic Alpha mining aims to identify simple formulas that can replace Alpha factors. Early approaches  centered on the manual extraction of factors with economic significance, such as the Fama three-factor model\cite{fama3}.  However, as these factors became widely known and used, their ability to predict outcomes started to decrease\cite{y7}. Over time, to overcome the limitations of traditional factors, researchers began exploring alternative methods, such as genetic programming, to automatically generate more effective factors. \cite{y7} conducts an analysis of 101 formulaic Alpha factors within the US market. Employing a genetic programming algorithm, the study systematically explores individual, unrelated factors through genetic variation of formula trees. The current  method involves using reinforcement learning algorithms to simultaneously identify a combination of Alpha factors along with their associated weights\cite{kdd2023}, with the objective of optimizing the discovery of the most robust composite factors. 

In investment practice, investment managers often collect a large batch of Alpha factors into a factor library. These alpha factors are combined through a combination model to form "Mega-Alpha", a final signal used for trading. With each new data arrival, factor values are recalculated, and Mega-Alpha is computed for trading decisions. Due to the financial market's emphasis on interpretability, the models for combining factors typically adopt linear structures. Cutting-edge Reinforcement Learning (RL) methods\cite{kdd2023} integrate the combination model and mining process into a unified framework, determining both factor discovery and their combination into Mega-Alpha. However, despite extensive factor exploration, only a fixed subset is typically utilized in actual investment. Additionally, the cyclic nature of each Alpha factor's stock selection capability, and the potential for reversal over time, must be considered. Fixed factors and weights sometimes render parts of the Mega-Alpha ineffective or even reverse their effects. 

To address the utilization rate of alpha factors and their fluctuations, we propose a two-stage alpha factor mining and combination framework. It consists of a generative-predictive neural network for mining factors and a composite model for dynamically selecting and combining factors based on their dynamic performance. 

The factor mining model, inspired by \cite{den}, utilizes a surrogate model to learn the distribution of alpha factor scores.  The generative model is trained to maximize the output of the surrogate model, facilitating the generation of high-scoring factors. Gradient-based methods ensure the generation of desirable factors even in the extremely sparse score space, allowing adjustments to the scoring function based on previously mined factors to ensure low correlation. 

The composite model forges dynamic Mega-Alpha from the alpha factors generated by the mining model. With each new trading day, it reassesses the performance of factors, selects those providing information for the next day's trading based on their performance, and calculates the optimal weights for the factor combination to produce that day's Mega-Alpha. This method considers the invalidation changes of factors and maximizes the use of Alpha factors produced by the mining model within a valid range, achieving a "mine as much as you use" efficiency. To evaluate our Alpha factor mining framework, we conducted a series of experiments. The results revealed that, compared to previous methods, our model can produce better Alpha factors and achieve higher returns in investment simulations. 
The main contributions of this paper are as follows:
\begin{enumerate}
    \item We introduce a generative-predictive factor mining model that leverages the powerful spatial exploration capabilities of deep learning to efficiently mine alpha factors even when the target function is sparse and complex. Moreover, the objective function in the factor mining process is variable.
    \item We propose a dynamic alpha factor combining model to generate Mega-Alpha. This approach enhances the traditional use of fixed-weight Mega-Alpha by allowing dynamic consideration of the time-varying effects of new market data, incorporating real-time dynamic weights.
    \item We conducted a comprehensive set of experiments to validate the effectiveness of our proposed methodology. Subsequent additional experiments and real investment provided evidence that factor timing can result in profitable outcomes.
\end{enumerate}

\section{Preliminary}
\subsection{Alpha Factor Definition}
In a market with \(n\) stocks, over \(T\) trading days where \(t \in \{1, 2, \ldots, T\}\), each stock is associated with a feature vector \(x_{ti} \in \mathbb{R}^{m\tau}\) on each trading day. The dataset \(X = \{X_t\}\) comprises \(m\) original features and rolling window data for the past \(\tau\) days. Moreover, for each stock on a given trading day, there exists a corresponding future return \(y_{ti} \in \mathbb{R}\), constituting the return matrix \(Y = \{y_t\}\). In this paper, six original features, namely open, high, close, low, volume, and vwap, are employed. An alpha factor \(f\) is defined as a function mapping the original feature matrix \(X_t \in \mathbb{R}^{n \times m\tau}\) of \(n\) stocks on a specific day to a factor value \(v_t = f(X) \in \mathbb{R}^n\).

\subsection{Alpha Factor Metrics}

The evaluation metrics includes IC, ICIR, Rank IC and Rank ICIR. The IC of factor $f$ represents the time-series average of Pearson's correlation coefficient between the factor value $v_t$ at time $t$ and the stock returns to be predicted $y_t$:
\begin{equation}
\label{eq1}
    IC(f, X, Y)=\mathbb{E}_{t}\left[\rho\left(v_{t}, y_{t}\right)\right]=\frac{1}{T} \sum_{t=1}^{T} \rho\left(v_{t}, y_{t}\right)
\end{equation}
where $v_t,y_t\in \mathbb{R}^n $.
The correlation $\rho\left(v_{t}, y_{t}\right)$ for each cross-section depicts the relationship between the factor value and the subsequent period's return. The IC describes the overall stock-picking ability of the factor, with higher values indicating superior stock-picking performance. In addition, the inclusion of RankIC is necessary to complement the measurement indicators because of the instability of pearson correlation. More details about these metrics could be found in the supplement materials.

\subsection{Formulaic Alpha}
The formalization of the Alpha Factor is represented by a mathematical expression formula. The formulaic operator $f$ is a function that maps the raw feature matrix $X_{t} \in \mathbb{R}^{n \times m \tau}$ of $n$ stocks on a given day to a factor value  $v_{t}=f(X) \in \mathbb{R}^{n}$ through a mathematical expression. 
The raw feature data of the $i-th$ stock on day $t$ is denoted as  $X_{t} \in \mathbb{R}^{n \times m \tau}$. The available data includes $m$ basic features for each of the  preceding $\tau$ days. The formula expression consists of operators and operands. The operands consist of $m$ basic features, along with optional constants. The operators encompass unary operators such as 'abs' and 'log,' as well as binary operators like +, -, *, /, and operators that account for time series considerations, such as Sum(\$volume,5d), indicating the summation of volume values over the past 5 days.

The representation of formulas traditionally involves expression trees, where operands are leaf nodes, and operators are non-leaf nodes. Initial attempts at formula generation employed genetic programming algorithms on these trees.  In alignment with contemporary deep learning methods, our approach follows the methodology introduced in \cite{kdd2023}. This involves representing formulas using Reverse Polish Notation (RPN), acquired through post-order traversal of the formula tree. This facilitates the representation of formulas in a one-hot matrix format compatible with deep learning networks.

\section{Methodology}
Our factor mining framework consists of two integral components:  (1) Alpha factor mining network employing a generative-predictive structure, wherein the Predictor serves as a surrogate model tasked with learning the distribution of alpha factor fitness, ie., the objective function. The Generator is trained to maximize the predicted values of the Predictor, thereby generating factors with elevated fitness. (2) A factor timing model, designed considering the time-series attributes of the factor. This model assigns weights to the factor, aiming to maximize the Information Coefficient (IC) of the Mega-Alpha formed by the combination of factor weights at each cross-section. The  proposed framework is illustrated in Figure \ref{onetwo}.
\subsection{Factor Mining Model}
  

\begin{algorithm}
\fontsize{6.5pt}{8pt}\selectfont
\caption{Factor Mining Pipeline}
\label{a1}
\begin{algorithmic}[0]
\Statex \textbf{Input:} stock data including data and target $X=\{X_t\}, Y=\{y_t\}$
\Statex \textbf{Output:} A group of respectively low correlation strong factors which is called factor zoo: $\mathcal{Z}=\{f_1,\ldots,f_k\}$
\Statex Initialize the factor zoo $\mathcal{Z}=\emptyset$
\Statex Sample a group of randomized factors onehot matrices $\mathcal{R}=\{\mathbf{x}_1,\ldots,\mathbf{x}_r\}$
\Statex \textbf{while} $|\mathcal{Z}| < \text{TargetFactorNum}$ \textbf{do}
\Statex \hspace{1em}$\mathcal{R}_{\text{fitness}}=\{\pi(\mathbf{x}_1,\mathcal{Z},X,Y),\ldots,\pi(\mathbf{x}_r,\mathcal{Z},X,Y)\}$
\Statex \hspace{1em}$\theta_G \gets \theta_G||\text{rand()}, \theta_P \gets \theta_P||\text{rand()}$
\Statex \hspace{1em}Train net $P$ with $\mathcal{R}$ and $\mathcal{R}_{\text{fitness}}$
\Statex \hspace{1em}\textbf{for} each epoch \textbf{do}
\Statex \hspace{2em}$\mathbf{z}_1, \mathbf{z}_2 \sim \mathcal{N}(0,1)^Q$
\Statex \hspace{2em}$\mathbf{x}_1 = M(G(\mathbf{z}_1)),\ \mathbf{x}_2 = M(G(\mathbf{z}_2))$
\Statex \hspace{2em}$\mathcal{L}(\theta_G) = \mathcal{L}_G(\mathbf{z}_1, \mathbf{z}_2, \mathbf{x}_1, \mathbf{x}_2, \theta_P)$
\Statex \hspace{2em}$\theta_G \gets \text{GradientDescent}(\mathcal{L}(\theta_G))$
\Statex \hspace{2em}$\mathcal{Z}_{\text{new}} = \text{parse}(\mathbf{x}_1) \cup \text{parse}(\mathbf{x}_2)$
\Statex \hspace{2em}\textbf{for} $f_{\text{new}}$ in $\mathcal{Z}_{\text{new}}$ \textbf{do}
\Statex \hspace{3em}\textbf{if} $f_{\text{new}}$ is qualified and $f_{\text{new}} \notin \mathcal{Z}$ \textbf{then}
\Statex \hspace{4em}$\mathcal{Z} \gets \mathcal{Z} \cup \{f_{\text{new}}\}$
\Statex \hspace{3em}\textbf{end if}
\Statex \hspace{2em}\textbf{end for}
\Statex \hspace{2em}$\mathcal{R} \gets \mathcal{R} \cup \{\mathbf{x}_1, \mathbf{x}_2\}$
\Statex \hspace{1em}\textbf{end for}
\Statex \textbf{end while}
\Statex \textbf{return} $\mathcal{Z}$
\Statex For demonstration convenience, the batch size dimension is not shown.
\end{algorithmic}
\end{algorithm}

\begin{figure*}[h]
  \centering
  \includegraphics[width=\linewidth]{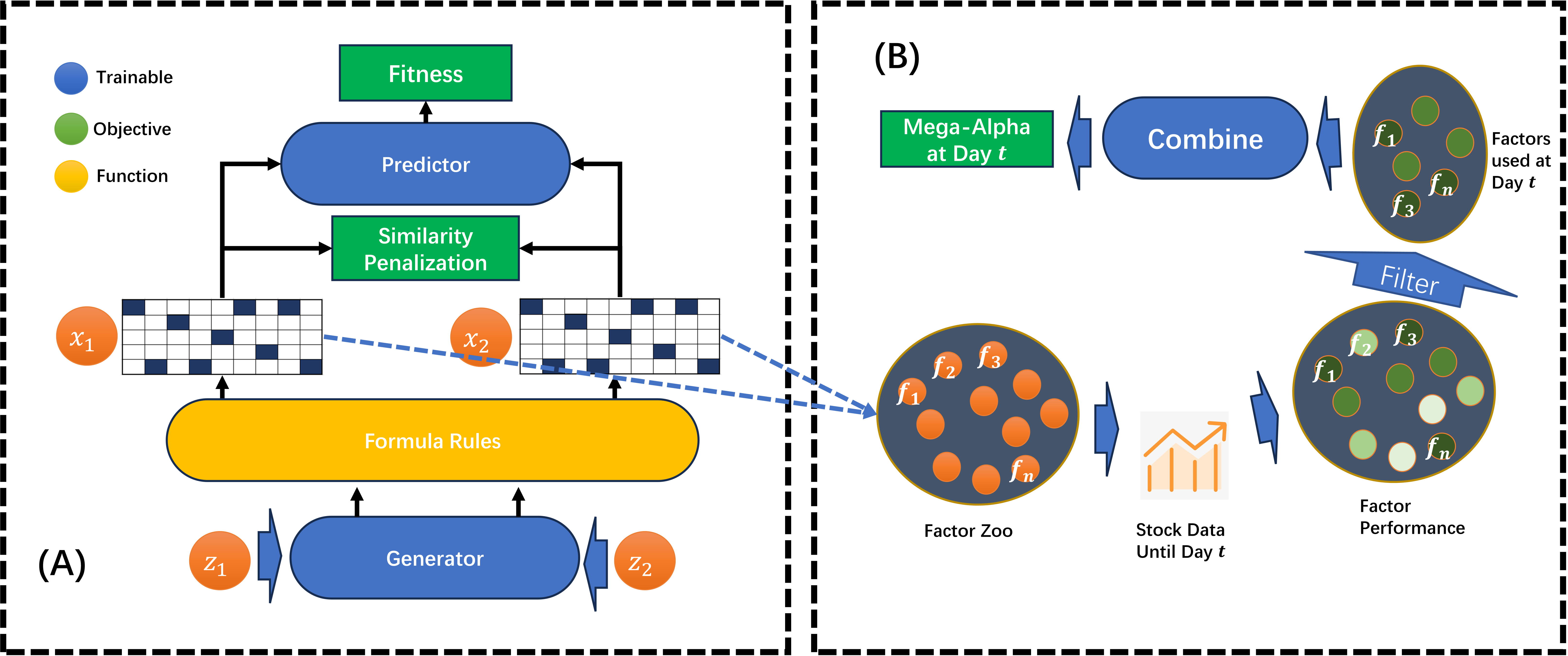}
  \caption{The illustration of our overall framework. (A) Alpha Factor Generating Model which generates the factor zoo. (B) Demonstrates the process of combining Mega-Alpha on day t, a process iteratively executed for each trading day.}
  \label{onetwo}
\end{figure*}

	Our factor mining model comprises a generator $G$ and a differentiable predictor $P$, as illustrated in part (A) of Figure \ref{onetwo}. The network $P(x)$ is dedicated to modeling the fitness score and undergoes training prior to the training of the network $G$, where $x_1 \in{0,1}^{D \times S}$ represents an one-hot matrix of an alpha factor. Here, $S$ denotes the maximum length of the formula, and $D$ denotes the number of all available operators and features. The training objective of $P$ is to predict the fitness score. The training process for the network $P$ is relatively straightforward, with training data sourced from the evaluation of all existing factor data in the sample library  
 $\mathcal{R}=\left\{x_{1}, \ldots, x_{r}\right\}$, and $\mathcal{R}_{\text {fitness}}=\left\{\right.$ fitness $\left(x_{1}\right), \ldots$, fitness $\left.\left(x_{r}\right)\right\}$.
 The training loss function is formulated as the mean squared error between the output of $P$ and the actual fitness score:
\begin{equation}
\fontsize{8pt}{0pt}\selectfont
\mathcal{L}_{P}=\sqrt{\frac{1}{n} \sum_{i=1}^{n}\left(P\left(\boldsymbol{x}_{i}\right)-\text { fitness }\left(\boldsymbol{x}_{\boldsymbol{i}}\right)\right)^{2}} .
\end{equation}
The generator network $G(z)$ takes a $Q$-dimensional normal distribution noise $z \in \mathbb{R}^{Q}$ as input. The output of $G(z)$ is a $D \times S$ logit matrix, which undergoes transformation into a one-hot matrix $\boldsymbol{x}=M(G(z)) \in{\{0,1\}}^{D \times S}$. This transformation involves the application of the operator $M()$ for sequence rule mask and gumbel-softmax. It is important to note that the $M()$ process maintains differentiability, allowing gradients to be propagated.

Once $P$ is trained, it serves as an estimation network for the formula fitness score. Subsequently, the parameters of $P$ are frozen, and the training objective shifts to maximizing the output of $P$. This training process involves training $G$ to generate formulaic alphas capable of maximizing the score of $P$.
\begin{equation}
\fontsize{8pt}{0pt}\selectfont
\mathcal{L}_{\text {Fitness }}=-P(M(G(z)))
\end{equation}
However, focusing solely on optimizing for high fitness may lead to premature convergence of the network $G$ to a local optimum. Therefore, it is necessary to introduce a diversity loss to force the generator $G$ to produce a diverse array of alpha factor formulas. To achieve this, we introduce a loss by generating two sets of factors based on two samplings of $z_1$ and $z_2$, subsequently penalizing the correlation between these two sets of factors. The final loss function for training the generator $\mathrm{G}$ is:
    

\begin{dmath}
\mathcal{L}_{G}=\mathcal{L}_{\text {Fitness }}+\mathcal{L}_{\text {Diversity }}=-P\left(\boldsymbol{x}_{1}\right)+\lambda_{\text {onehot }} * \text { Similarity }_{\text {onehot }}\left(f\left(\mathbf{z}_{1}\right), f\left(\mathbf{z}_{2}\right)\right)+\lambda_{\text {hidden }} * \text { Similarity }_{\text {hidden }}\left(f\left(\mathbf{z}_{1}\right), f\left(\mathbf{z}_{2}\right)\right)
\end{dmath}

Our framework is designed to identify a set of highly effective strong factors in the factor mining stage for the factor timing combination model. This involves an investigation, including a comparison between strong and weak factor pools. A strict criterion is maintained to ensure that factors included in the library meet specific requirements. Incorporating domain knowledge, our criteria for factor entry comprise three fundamental aspects:IC and ICIR are used to filter the stock-picking ability and stability, the correlation of returns with existing factors in the library avoids overlapping with the stock-picking ability of existing factors.
Due to the implementation of a generative-predictive architecture coupled with gradient-based algorithms, the generator is capable of capturing the essential characteristics of factors in a "directional" manner, even under conditions of significant sparsity within the fitness function:
\begin{dmath}
\small
\pi(x, Z, X, Y)=
\left\{\begin{aligned}
\text { Abs }(IC(f, X, Y)),   \; & f \text { is valid and  }\psi(f, \mathcal{Z}, X, Y)<\operatorname{CORR}^{\prime} \\
0, \; & \text { else }
\end{aligned}\right.
\end{dmath}
Where $f=\operatorname{parse}(x)$  represents the operational formula parsed from the onehot matrix representation $x$, and $\mathcal{Z}$ denotes the existing factor set which is called factor zoo. The absolute value of IC is taken because a negative IC factor can be transformed into a positive IC factor by reversing the value of the factor. The $\psi$ function  computes the maximum absolute value of the correlation between $f$ and each existing factor in $\mathcal{Z}$. Here, CORR' is a manually set parameter. When $|Z|=0$, $\pi(f, \mathcal{Z}, X, Y)$ returns the absolute value of the factor's IC. Algorithm \ref{a1} provides a detailed description of our alpha mining model.

\subsection{Alpha Combination}

\begin{algorithm}
\fontsize{6.5pt}{8pt}\selectfont
\caption{Factor Combining Pipeline}
\label{algo2}
\begin{algorithmic}
\Statex \textbf{Input:}Factor zoo $\mathcal{Z}=\{f_1,\ldots,f_k\}$, max factor number $N$, dataset $X=\{X_t\}, Y=\{y_t\}$
\Statex \textbf{Output:} Prediction  $\hat{Y}=\{\hat{y_t }\}$
\Statex $\hat{Y}\xleftarrow{}{\phi}$
\Statex \textbf{for} $t \gets 1$ \textbf{to} $T$ \textbf{do}
\Statex \hspace{1em}$\mathcal{Z}_t={\phi}$
\Statex \hspace{1em}\textbf{for all} $f \in \mathcal{Z}$ \textbf{do}
\Statex \hspace{2em}Calculate $IC_t \rho(f),ICIR_t \hat{\rho}(f)$ 
\Statex \hspace{2em}\textbf{if} $IC_t \rho(f)>IC^\prime{}$  $\And$  $ICIR_t \hat{\rho}(f)>ICIR^\prime$ \textbf{then}
\Statex \hspace{3em}$\mathcal{Z}_t\xleftarrow[]{}\mathcal{Z}_t \cup\{f\}$
\Statex \hspace{2em}\textbf{end if}
\Statex \hspace{1em}\textbf{end for}
\Statex \hspace{1em}Sort $\mathcal{Z}_t$ based on $IC_t \rho(f)$ 
\Statex \hspace{1em}Select the top $N$ factors from $\mathcal{Z}_t$: $\mathcal{Z}_t^{(N)} = \text{Top-N}(\mathcal{Z}_t)$
\Statex \hspace{1em}$Model=LinearRegression(\mathcal{Z}_t^{(N)},y_t)$
\Statex \hspace{1em}$\hat{y_t}\xleftarrow[]{}Model. Predict(X_t)$
\Statex \hspace{1em}$\hat{Y}\xleftarrow[]{}\hat{Y} \cup \hat{y_t}$
\Statex \textbf{end for}
\Statex \textbf{return} $\hat{Y}$

\end{algorithmic}
\end{algorithm}

In the investment process, there exists a significant demand for interpretability. Investors commonly find it challenging to accept a model that operates as an inexplicable black box. A qualified investment manager is required to possess an understanding of the factors influencing portfolio performance.  This involves exploring the logic behind factors` effects, an assessment of factors prone to failure or change, and the necessity for regular adjustments to the factors incorporated into the final model. Additionally, nonlinear combination models are susceptible to overfitting in financial datasets. Consequently, linear models are typically favored as the primary choice due to their ability to mitigate concerns related to overfitting.

Given the potential periodic or permanent ineffectiveness of certain factors due to congestion, shifts in market style, etc., the utilization of fixed factor weights proves inadequate in promptly adjusting to changes in factor weights. This inadequacy can result in overfitting of the training set.  The factor metrics serve as performance indicators over a specific time frame. Upon the arrival of new data, the evaluation indicators for factors undergo alterations. Due to the momentum effect observed in factor performance, factors that have demonstrated success in the past tend to exhibit positive performance in the future\cite{momentum}. 

To address these challenges, we have developed a dynamic weight factor combination model. At each time point $t$, leveraging data from the preceding $n$ days, we conduct a reassessment of the factors within the factor zoo $\mathcal{Z}$. The factors are re-ranked and selected based on their most recent performance metrics, including ICIR, IC, RankIC, etc. Subsequently, we employ the latest data to fit the best linear model for predicting the current combination of `N' factors. This model is then utilized to predict the current data point. The detailed workings of our combinational model are demonstrated in Algorithm \ref{algo2}  and (B) at Figure \ref{onetwo} .

	The combination algorithm we have developed demonstrates the ability to adjust promptly the components and composition weights of the final Meta-Alpha in accordance with the performance of the factors. This intuitive adaptability enhances its effectiveness in responding to market changes while simultaneously upholding the imperative of maintaining explainability. These observations are validated by the outcomes of our experiments.

\subsection{Overview}

The over all AlphaForge framework is shown in Figure \ref{onetwo}. The initial step involves training a generative model using the training set data, aiming to maximize the Information Coefficient (IC). This process aims to generate a batch of alpha factors characterized by low correlation and high quality, meeting predefined criteria and encompassing a diverse range of price-related information. Subsequently, this collection of factors is archived into a repository referred to as the Factor Zoo.

Once the Alpha Factor Zoo has been extracted,it serves as a fixed input to the combination model and remains unchanged thereafter. During the inference and trading phases, the combination model utilizes updated historical data at each time step $t$ to reassess the recent performance of each factor within the Factor Zoo. Based on this evaluation, the model filters and integrates factors to formulate the Mega-Alpha signal for the given day. The hyperparameters could be found in supplementary materials and our framework implementation is published on GitHub\footnote{\url{https://github.com/DulyHao/AlphaForge}}.

\section{Experiments}

The purpose of our experimental design is to answer the following questions: \newline 
$\textbf{Q1}$: Does our framework outperform the previous formula-based Alpha factor approaches? 
\newline $\textbf{Q2}$: How does the performance of our model vary with changes in the pool size of the factors? 
\newline $\textbf{Q3}$: Is each component of our model framework effective?
\newline $\textbf{Q4}$: How does our framework perform in real-world trading scenarios?

\subsection{Experiments settings}
\subsubsection{Data}
We choose the CSI300 and CSI500 dataset because the constituent stocks of these two indices cover the majority of the market capitalization in China's A-share market. Additionally, the studies we referenced also focused on these two datasets\cite{kdd2023,y23,y25}. The market styles are diverse and ever-changing, potentially leading to over-fitting of models during the training, validation, and testing dataset splits. In practical investment contexts, over-fitting often yields adverse outcomes. Moreover, in real-world investment practices, it is essential to periodically re-calibrate models with the influx of new real-time data. To mitigate over-fitting and closely emulate the actual investment process, we conducted performance testing from 2018 to 2022. The model was retrained annually with updated data, using the year preceding the test year as the validation dataset, resulting in a total of five training sessions. The first training set, validation set and test set are respectively (2010-01-01 to 2016-12-31), (2017-01-01 to 2017-12-31) and (2018-01-01 to 2018-12-31). We use 'Ref(VWAP, -21)/Ref(VWAP, -1) - 1' as the label because it more closely reflects real-world scenarios, although it may lead to changes in the metrics. More details on this can be found in the supplementary materials.  The stock data is public data and we use Qlib\cite{y25} to download it.


\subsubsection{Compared Methods}
To assess the distinction of our framework in comparison to traditional formulaic Alpha factor generation methods, we designed three methodologies for comparison against our approach. These include the Genetic Programming (\textbf{GP}) method, the Deep Symbolic Optimization \textbf{DSO} method \cite{DSO}  and Reinforcement Learning (\textbf{RL})\cite{kdd2023}. \textbf{GP} employs the Information Coefficient (IC) as the optimization objective, generating formula trees through genetic approaches. As a representative method of symbolic regression, the\textbf{ DSO} method aligns best with our task. \textbf{RL} utilizes reinforcement learning techniques to generate a set of alpha factors, with the optimization objective being the IC of a Mega-Alpha composed of these factors. To avoid the effects of random seeds, we repeated the run 5 times for each model.

We further incorporated three Machine Learning based models for benchmarking purposes: \textbf{XGBoost}\cite{y2}, an ensemble learning method using gradient boosting with decision trees, effective for capturing non-linear relationships in stock market data; \textbf{LightGBM}\cite{y8}, a highly efficient gradient boosting framework, optimized for handling large-scale financial datasets; and \textbf{MLP (Multilayer Perceptron)}, a neural network model capable of learning complex patterns in financial data through multiple interconnected layers of nodes.

\subsubsection{Additional Experiments}
To answer $\textbf{Q2}$ we conducted experimental comparisons for different alpha pool size limitations set at [1, 10, 20, 50, 100]. To address $\textbf{Q3}$ and demonstrate the efficacy of different components within our model, we removed the dynamic combination method from our model and conducted comparative experiments against the complete model.  For $\textbf{Q4}$, we carried out real world data simulation trading experiments over a continuous five-year dataset for comparison. 

Additionally, due to space constraints, the user survey conducted among industry professionals and other supplementary experiments can be found in the supplementary materials.

\subsection{Main Results}
Regarding $\textbf{Q1}$: As shown in Table \ref{table1}, our method demonstrates superior performance across various metrics, including stock selection ability indicators such as IC and RankIC. We conduct comparative experiments with non-formulative MLP,LlghtGBM, XGBoost, as well as with formulative methods GP, RL and DSO. Table 1 shows that our method outperforms all the methods in CSI300 and CSI500 datasets. This indicates that the AlphaForge framework has achieved notable advancements in stock selection ability compared to the baseline methods. 

\begin{table}[h]
\fontsize{9}{9}\selectfont 
\setlength{\tabcolsep}{1mm}
\small
\centering
\caption{Comparison of Methods on CSI 300 and CSI 500}
\begin{tabular}{lcc cc}
\textbf{} & \multicolumn{2}{c|}{\textbf{CSI 300}} & \multicolumn{2}{|c}{\textbf{CSI 500}} \\ \hline
 & \textbf{IC}(\%) & \textbf{RankIC}(\%) & \textbf{IC}(\%) & \textbf{RankIC}(\%) \\ \hline
XGB        & 0.41 & 1.63 & 0.33 & 2.87 \\ 
MLP            & 1.22(0.16) & 1.75(0.28) & 1.94(0.11) & 3.31(0.23) \\ 
LGBM       & 0.84 & 1.85 & 1.75 & 3.81  \\ 
               \hline
GP      & 1.29(0.44) & 2.72(0.58) & 0.37(0.76) & 2.34(1.07) \\ 
DSO      & 2.55(0.69) & 3.88(1.12)  & 1.38(0.57) & 4.56(0.61) \\ 
RL      & 2.09(0.26) & 2.72(0.42) & 1.91(0.49) & 4.03(0.62) \\  \hline
Static    & 2.43 (0.57)& 3.67(0.46) & 2.05(0.29) & 4.48(0.46) \\ 
\textbf{Ours}& \textbf{4.40(0.56)} & \textbf{5.89(0.69)} & \textbf{2.84(0.58)} & \textbf{5.57(0.58)} \\ \hline
\end{tabular}
\label{table1}
\end{table}

\subsection{Effect of the Pool Size}

\begin{figure}[h]
  \centering
  \includegraphics[width=\linewidth]{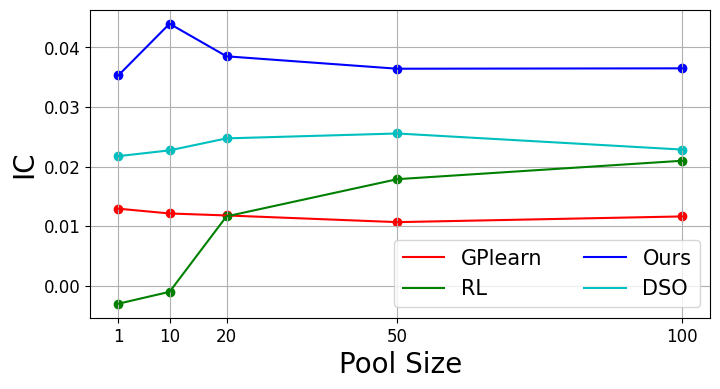}
  \caption{The IC in CSI300 across Different Pool Size}
  \label{fig3}
\end{figure}
To answer \textbf{Q2}, we varied the size of the Alpha factor pool to [1,10,20,50,100], respectively, to examine the influence of factor pool size on performance. As our model dynamically determines factor weights, the composition of "Mega-Alpha" can vary over time, with the total number of factors limited to not exceed the pool size. The results in Figure \ref{fig3} reveal a non-monotonic relationship between our method's performance and the factor pool size, with the highest performance observed when the pool size is set to 10. We attribute this phenomenon to the dynamic selection of factors by the combination model. Not all factors are consistently effective, and at any given time, approximately 10 factors capture the most relevant price information. Thus, further increasing the factor library size could potentially yield diminishing returns for "Mega-Alpha."
\subsection{Ablation Study}
In addressing $\textbf{Q3}$, we conducted experiments by selectively excluding certain components. Table \ref{table1} presents the results. Specifically, "Static" refers to the utilization of our alpha factor mining model for generating alpha and composing the Mega-Alpha using the same approach as "RL". Conversely, "Dynamic" involves employing our full model version. Our findings indicate that our predictive-generative alpha factor mining method achieves superior results compared to the previous state-of-the-art algorithm. Furthermore, the superior performance of the "Dynamic" model over the "Static" model confirms the efficacy of our dynamic factor combining approach.

\subsection{Case Study}

\begin{table}[htbp]
  \centering
  \caption{Factors used on Day 1}
  \label{tab1}
  {
  \fontsize{7}{8}\selectfont 
  \begin{tabular}{c|p{6cm}|c}
    \toprule
    \textbf{\#} & \textbf{exprs} & \textbf{weight} \\ \hline
    \midrule
    1 & S\_log1p(ts\_cov(high,volume,20)) & -0.00092\\
    2 & S\_log1p(ts\_min(ts\_corr(high,volume,5),10)) & -0.00180\\
    3 & S\_log1p((-10.0-ts\_corr((close+0.01),(0.5+volume),30))) & -0.00014\\
    4 & S\_log1p(ts\_min(ts\_cov(high,volume,5),1)) & -0.00178\\
    5 & S\_log1p(ts\_min(ts\_corr(close,volume,10),1)) & -0.00029\\
    10 & (Inv((Inv(S\_log1p(ts\_mad((S\_log1p(ts\_corr(high,volume,10))*\newline Inv((S\_log1p(volume)-30.0))),20)))/30.0))+2.0)& 0.00171\\
    32 & Inv((((ts\_cov(vwap,(((-30.0-S\_log1p((volume/-2.0)))+-10.0)*10.0),30)+5.0)/5.0)-30.0)) & 0.00174\\
    36 & ts\_cov(close,volume,10) & -0.00031\\
    45 & ts\_std((Inv((-2.0-ts\_mad(S\_log1p(volume),50)))*2.0),40) & -0.00145\\
    54 & (S\_log1p(((-30.0+(S\_log1p(ts\_std(S\_log1p((volume*-10.0)),40))/-0.01))*2.0))--30.0) & 0.00132\\
    \bottomrule
  \end{tabular}
  }
\end{table}

\begin{table}[htbp]
  \centering
  \caption{Factors used on Day 2 }
  \label{tab2}
    {
  \fontsize{7}{8}\selectfont 
  \begin{tabular}{c|p{6.2cm}|c}
    \toprule
    \textbf{\#} & \textbf{exprs} & \textbf{weight} \\  \hline
    \midrule
    2 & S\_log1p(ts\_min(ts\_corr(high,volume,5),10)) & -0.00239\\
    3 & S\_log1p((-10.0-ts\_corr((close+0.01),(0.5+volume),30)))& 0.00168\\
    6 & (((30.0-ts\_mad(Ref(ts\_delta(ts\_corr(volume,vwap,10),1),10),50))--10.0)+-1.0) & -0.00200\\
    36 & ts\_cov(close,volume,10) & -0.00143\\
    45 & ts\_std((Inv((-2.0-ts\_mad(S\_log1p(volume),50)))*2.0),40) & -0.00040\\
    46 & ((((10.0-ts\_min(((ts\_corr(volume,(close/-0.01),40)**10.0)*-5.0),20))--10.0)*10.0)-5.0) & -0.00020\\
    54 & (S\_log1p(((-30.0+(S\_log1p(ts\_std(S\_log1p((volume*-10.0)),40))/-0.01))*2.0))--30.0) & 0.00167\\
    62 & Inv(((((ts\_mad((30.0*(S\_log1p(ts\_var(S\_log1p(volume),50))*5.0)),\newline 20)+2.0)--0.01)+-1.0)--0.01))& -0.00018\\
    63 & (Inv(Inv((S\_log1p(ts\_std((30.0*(S\_log1p(ts\_std(S\_log1p(volume)\newline ,50))*-0.01)),20))-0.5)))-10.0)& -0.00148\\
    99 & (((Inv(((30.0*(S\_log1p(ts\_std(S\_log1p(volume),50))*5.0))-0.01))-0.5)+30.0)-5.0)& 0.00127\\
    \bottomrule
  \end{tabular}
}
\end{table}

A case study of a composite model was extracted to illustrate our framework's capability in dynamic factor timing. Our generative model produced a factor zoo with 100 alpha factors, while the factor pool limit of the composite model was set at 10. Tables \ref{tab1} and \ref{tab2} present the composition of the "Mega-Alpha" factor and the corresponding weights on two distinct trading days, respectively. It is observed that, among the 10 alpha factors selected on the first trading day, only 5 are used on the second trading day. Notably, Factor 3 had a weight of -0.00014 on the first trading day, whereas its weight shifted to 0.00168 on the second trading day. This indicates that the same factor contributed differently to the "Mega-Alpha" generated by our model on different dates, highlighting the effectiveness of our framework in leveraging diverse Alpha factors to generate signals on various dates. This underscores a process of timing selection for alpha factors within our framework.

\subsection{Interpret-ability of Alpha Factors}
Taking factor 1 in table \ref{tab1} as an example, this factor can be interpreted as whether the trend of the high price and the volume are the same in the past 20 days. The negative weight of this factor reflects an underlying investment logic: When prices are rising but attract little attention, it may be worth considering a purchase. When prices are dropping and the crowd is panic selling, it may present an opportunity to buy. 

Another example is the factor `-1*ts\_mean(volume,20)', which represents the opposite of average trading volume over the past 20 days. This factor has a strong correlation with the stock’s market capitalization. If the weight of this factor is too large in the model, it can cause the portfolio to lean towards small-cap stocks, leading to dangerous risk exposure. Usually, a competent investment manager should seek to reduce the weight of this factor or use other methods to avoid excessive exposure to small-cap stocks.

Investment managers could analyze the inherent meaning and return sources of each factor in the model and make adjustments based on their understanding of the market. Interpretability helps the investment manager to analyze the sources of returns and make adjustments and attributions in real-world investments based on their market insights.

\subsection{Simulated Trading and Real Money Investment}
To assess the practical efficacy of our model, we conducted simulated trading based on the prediction results. The simulation trading period spanned from January 1, 2018, to December 31, 2022, utilizing the CSI300 stock pool. We employed the Qlib\cite{y25} framework, where the trading strategy entailed holding the top 50 stocks with the highest Mega-Alpha scores on a daily equal-weighted basis. Additionally, a daily limit of changing a maximum of 5 stocks was imposed to avoid excessive trading costs.
\begin{figure}[h]
  \centering
  \includegraphics[width=\linewidth]{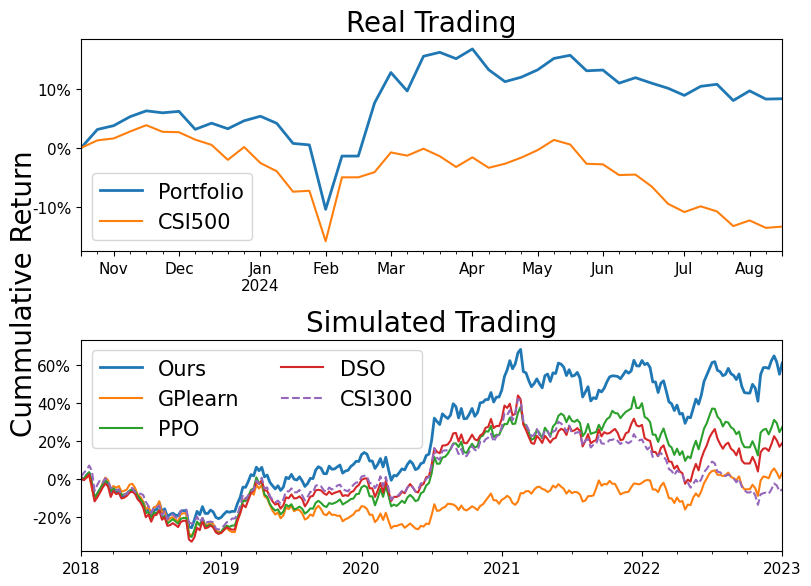}
  \caption{The Real(top) and simulated(bottom) Trading Result}
\label{f2}
\end{figure}

The top half shows our actual trading results. A real account was used with an investment of 3 million RMB in CSI500. Untill now, after approximately 9 months of investment, it has generated a 21.68\% higher excess return compared to CSI500.
By tracking the net value of funds in a simulated trading account. The bottom half of  Figure \ref{f2} illustrates the cumulative returns of different algorithms. The results show that our framework is able to achieve the best net account value in a simulation trade lasting for five years. It performs the strongest among all the comparative models. 

\section{Related Work}
\textbf{Formulaic Alpha Factors}: The exploration of Alpha factor formulation encompasses a vast search space. Genetic programming algorithm has historically been employed to generate factors through tree-based genetic mutations. Early advancements, notably within the GPLearn package,\cite{y10} introduced time series operators, establishing the first genetic programming method for mining alpha factors . \cite{y9} employed mutual information as the objective for factor mining to discover factors based on nonlinear relationships. \cite{y27} utilized the IC between Alphas to filter overly similar Alphas, enhancing the diversity of mined Alpha factors. AlphaEvolv \cite{y3} aimed at improving existing Alpha factors. Presently, \cite{kdd2023} is based on reinforcement learning (RL) for synergistic Alpha factor mining, offering a novel approach beyond genetic programming.  This method leverages the space exploration capability of reinforcement learning to mine a set of collaborative Alphas. However, current methods often fail to account for time-varying factor effects, typically adopting fixed factor combination weights and immutable objective functions during mining.

\textbf{Machine Learning-based Alpha Factors}: Methods for predicting stock returns using deep learning are also thriving. Early approaches did not consider interactions between stocks, merely using the historical time series data of each stock to predict stock prices, including Multilayer Perceptron (MLP), Transformer \cite{y21}, LSTM \cite{y5},  as well as tree-based methods like LightGBM \cite{y8} and XGBoost \cite{y2}. Subsequent developments led to models specifically designed for this task, such as HIST \cite{y23} which enhances traditional time series models by incorporating industry and concept graph data to capture the correlation information between different stocks. FactorVAE\cite{factorvae} combines the dynamic factor model (DFM) and variational autoencoder (VAE) to estimate the variance of the latent space distribution while predicting stock returns. These machine learning-based methods, when compared to the Formulaic Alpha approach, often lack interpret-ability.

\textbf{Symbolic regression} addresses the problem of identifying relationships between different variables, typically aiming to derive a single interpretable mathematical expression to solve scientific problems. As one of the few interpretable machine learning methods, symbolic regression finds extensive applications in fields such as mathematics, physical equations, and materials science. Early approaches to symbolic regression focused on improving genetic programming (GP). For instance, \cite{EPLEX} used lexicase selection for regression, \cite{AFP} introduced the age-layered population structure, and \cite{GSGP} employed semantic variation operators to generate offspring. With the advancement of neural network technology, \cite{champion2019data} proposed a new method by combining autoencoder networks with symbolic regression. \cite{biggio2021neural} emphasized the role of large-scale pre-training based on the transformer model, and \cite{DSO} introduced a unified framework of neural networks and symbolic regression.

\textbf{Discussion}: Our framework introduces a novel approach for mining formulaic Alpha factors, with a primary focus on the stock investment domain. This framework could be expanded to many other domains including but not limited to traffic flow prediction\cite{traffic}, sales forecasting\cite{sale1,sale2}, and customer churn prediction\cite{customer}. Despite the optimization objective of our mining model changing with the increase of the factor zoo, the training objective of our framework's mining algorithm's generator is the IC of individual alpha factors. Exploring the possibility of incorporating the combined IC of the entire batch of factors may enhance the efficiency of factor mining.
\section{Conclusion}
This paper introduces a new framework, named AlphaForge, for mining and dynamically combining formulaic alpha factors, thereby providing new tools for investors engaged in quantitative trading.  Our AlphaForge framework is able to leverage the powerful space exploration capabilities of deep learning models to explore the search space for formulaic Alphas. The design of the surrogate model to predict Fitness Score ensures that our model can efficiently generate Alpha factors using gradient methods, even in scenarios characterized by a sparse search space. We also introduce a composite model capable of dynamically combining factor weights at each time slice, allowing the Mega-Alpha generated by the model to timely adjust its component factor and their weights to adapt to market fluctuations. Through extensive experiments, we have demonstrated that our framework can achieve better performance compared to previous methods of formulaic alpha models. In more realistic trading tests, our model consistently delivers higher profits and has already earning us excess returns during real money trading.

\appendix

\bigskip
\nocite{*}
\bibliography{aaai25}

\section{Check List}

This paper:

Includes a conceptual outline and/or pseudocode description of AI methods introduced (yes)

Clearly delineates statements that are opinions, hypothesis, and speculation from objective facts and results (yes)

Provides well marked pedagogical references for less-familiare readers to gain background necessary to replicate the paper (yes)

Does this paper make theoretical contributions? (no)

Does this paper rely on one or more datasets? (yes)

If yes, please complete the list below.

A motivation is given for why the experiments are conducted on the selected datasets (yes)

All novel datasets introduced in this paper are included in a data appendix. (NA)

All novel datasets introduced in this paper will be made publicly available upon publication of the paper with a license that allows free usage for research purposes. (NA)

All datasets drawn from the existing literature (potentially including authors’ own previously published work) are accompanied by appropriate citations. (yes)

All datasets drawn from the existing literature (potentially including authors’ own previously published work) are publicly available. (yes)

All datasets that are not publicly available are described in detail, with explanation why publicly available alternatives are not scientifically satisficing. (NA)

Does this paper include computational experiments? (yes)

If yes, please complete the list below.

Any code required for pre-processing data is included in the appendix. (yes).

All source code required for conducting and analyzing the experiments is included in a code appendix. (yes)

All source code required for conducting and analyzing the experiments will be made publicly available upon publication of the paper with a license that allows free usage for research purposes. (yes)

All source code implementing new methods have comments detailing the implementation, with references to the paper where each step comes from (yes)

If an algorithm depends on randomness, then the method used for setting seeds is described in a way sufficient to allow replication of results. (yes)

This paper specifies the computing infrastructure used for running experiments (hardware and software), including GPU/CPU models; amount of memory; operating system; names and versions of relevant software libraries and frameworks. (yes)

This paper formally describes evaluation metrics used and explains the motivation for choosing these metrics. (yes)

This paper states the number of algorithm runs used to compute each reported result. (yes)

Analysis of experiments goes beyond single-dimensional summaries of performance (e.g., average; median) to include measures of variation, confidence, or other distributional information. (yes)

The significance of any improvement or decrease in performance is judged using appropriate statistical tests (e.g., Wilcoxon signed-rank). (yes)

This paper lists all final (hyper-)parameters used for each model/algorithm in the paper’s experiments. (yes)

This paper states the number and range of values tried per (hyper-) parameter during development of the paper, along with the criterion used for selecting the final parameter setting. (yes)

\end{document}